\RequirePackage{lineno}
\documentclass{PoS}

\usepackage{amsmath}
\usepackage{float}
\usepackage[square,numbers]{natbib}

\title{Studies of Heavy Flavor Jets Using $D^{0}$-hadron Correlations in Azimuth and Pseudorapidity in Au+Au Collisons at 200 GeV at the STAR Experiment}

\ShortTitle{$D^0$-hadron Correlations in Heavy Ion Collisions}

\author{\speaker{Alexander M. Jentsch (for the STAR Collaboration)}\\
        The University of Texas at Austin, Texas, USA\\
        E-mail: \email{alex.jentsch@utexas.edu}}

\abstract{Heavy flavor (HF) quarks (charm, bottom) are important probes of the medium produced in relativistic heavy-ion collisions because they are formed in the early stage and propagate throughout the lifetime of the QGP medium. HF-meson spectra and azimuthal anisotropy ($v_{2}$) measurements have been reported by experiments at RHIC and the LHC, and they suggest strong interactions of HF quarks with the medium. $D^{0}$-hadron correlations on relative pseudorapidity and azimuth ($\Delta\eta,\Delta\phi$) provide a method for disentangling correlation structures on ($\Delta\eta,\Delta\phi$),  related to jets and bulk phenomena directly, with the $D^{0}$ serving as a proxy for a charm jet. In these proceedings, we present 2D $D^{0}$-hadron angular correlations as a function of centrality in Au+Au collisions at $\sqrt{s_{NN}}$ = 200 GeV. These data reveal a jet-like, peaked structure at ($\Delta\eta,\Delta\phi$) = (0, 0) (near-side), and a $\Delta\eta$-independent azimuthal harmonic modulation. Here, we focus on the evolution of the near-side peak's yield and widths on ($\Delta\eta,\Delta\phi$) as a function of centrality and compare them to results from light flavor correlations with a similar trigger mean-$p_{T}$.}

\FullConference{International Conference on Hard and Electromagnetic Probes of High-Energy Nuclear Collisions\\
		30 September - 5 October 2018\\
		Aix-Les-Bains, Savoie, France}

\begin{document}


\section{Introduction}

Heavy flavor (HF) quarks are formed in the early stage of the collision between ultra-relativistic heavy ions via hard partonic scatterings. HF quarks rapidly hadronize, and then decay outside the deconfined partonic medium, or quark-gluon plasma (QGP). Since HF quarks are formed early and decay late, they can sample the entire evolution of the QGP and are therefore ideal probes for the study of the complicated QGP medium.  

Recently, measurements of the nuclear modification factor, $R_{AA}$, by the STAR and ALICE collaborations show significant differences between light flavor (LF) mesons and HF mesons at low transverse momentum for mid- to very-central heavy-ion collisions \cite{STAR_HADRON_RAA_PAPER,STAR_D0_RAA_PAPER, ALICE_D_MESON_RAA_PAPER}. However, at higher values of $p_{T}$, the $R_{AA}$ of the LF and HF mesons are consistent with each other. Furthermore, the STAR collaboration has measured the $D^{0}$-meson azimuthal anisotropy, $v_{2}$, and seen that it is consistent with the $v_{2}$ of LF mesons as a function of transverse kinetic energy ($m_{T}-m_{0}$) at 10-40\% centrality \cite{STAR_D0_V2_PAPER}. While no measurements can currently disentangle the effects of collisional and radiative energy loss, two-particle correlations can yield additional information from the net effects of interactions in the QGP.

Two-particle correlations allow for the study of the complicated underlying dynamics of heavy-ion collisions. In particular, correlations on both relative azimuth ($\Delta\phi$) and pseudorapidity ($\Delta\eta$) using a $D^{0}$ meson as a trigger provide access to both jet-like physics - with the $D^{0}$ serving as a proxy for a charm jet - and to flow harmonics dependent on $\Delta\phi$, which allows for simultaneous extraction of the evolution of the jet-like peak, as well as $v_{2}$, using a multi-parameter fit. In these proceedings, we focus on correlations at small $\Delta\phi$ ($|\Delta\phi| \ < \ \pi/2$; near-side or NS) and $\Delta\eta$, which provide access to the charm jet interactions with the medium, and study the evolution of the NS correlation peak as a function of the centrality of the collision.

\section{$D^{0}$-hadron Correlations}
\subsection{Dataset and Trigger/Associated Reconstruction Cuts}
The analysis presented in these proceedings was carried out using around 900 million minimum-bias events collected by the STAR detector \cite{STAR_DETECTOR} at the Relativistic Heavy-Ion Collider (RHIC) in 2014. Particle trajectories (tracks) were reconstructed using the Time Projection Chamber (TPC) \cite{STAR_TPC} and the Heavy Flavor Tracker (HFT) \cite{STAR_HFT}, where the TPC is used for extraction of particle momenta. The HFT is used for precise calculation of secondary decay vertices displaced from the primary collision vertex (PV) by $\sim$ 100 $\mu$m, enabling reconstruction of HF hadrons via rejection of combinatorial background coming from particles originating at the PV.

The TPC and HFT both have full 2$\pi$ coverage in azimuth, and both have pseudorapidity coverage of $|\eta| \ < \ 1$. The $D^{0}$-meson was reconstructed via the hadronic decay channel ($D^{0}(\bar{D}^{0}) \rightarrow K^{\mp} \ + \ \pi^{\pm}$), with $2 \ < \ p_{T, D^{0}} \ < 10$ GeV/$c$, with the range based on experimental constraints and statistics. The reconstruction of the $D^{0}$ uses five topological cuts mostly based on the distance of closest approach (DCA) of tracks, which are detailed in Fig. \ref{D0_Decay_Cartoon}. The associated hadron tracks used in this analysis require hits in both the TPC and the HFT detectors with acceptance cuts $|\eta| \ < \ 1$ and $p_{T} \ > \ 0.15$ GeV/$c$.

\subsection{Calculation of Correlations}

\begin{figure}[H]
\centering
\includegraphics[width=5 cm]{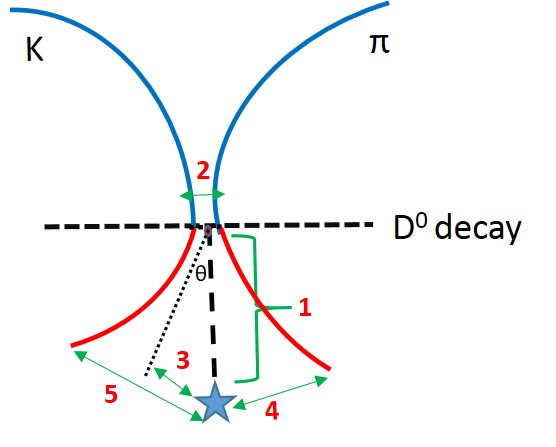}
\includegraphics[width=5 cm]{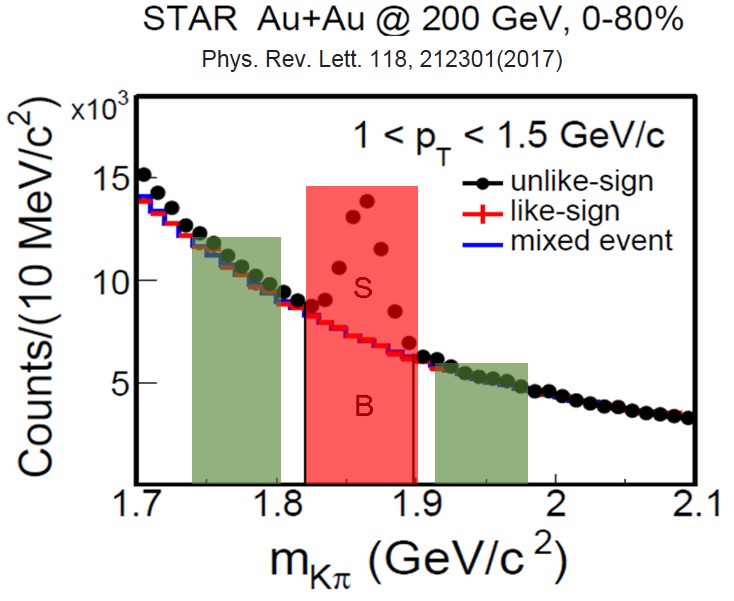}
\caption{Left panel: Cartoon sketch of the $D^{0}$ decay to an unlike-sign K$\pi$ pair. The decay components are enumerated as follows: 1) decay length of the mother $D^{0}$, 2) DCA K$\pi$ daughters, 3) DCA of $D^{0}$ to PV, 4) DCA of $\pi$ to PV, and 5) DCA K to PV. Right panel: $D^{0}$ invariant mass distributions with the signal region highlighted in red, and the sidebands highlighted in green. }
\label{D0_Decay_Cartoon}
\end{figure}

After reconstruction of the candidate $D^{0}$s and selection of the associated hadrons, the K$\pi$-hadron pairs from the same-event (SE) are binned on ($\Delta\eta, \Delta\phi$), with each pair being weighted by an efficiency correction factor. In order to correct for detector acceptance effects, $D^{0}$ candidates are paired with associated hadrons from different events, or \textit{mixed-events} (ME),  with a similar event multiplicity and z-coordinate of the PV. The correlation is based on Pearson's correlation coefficient, as seen in \cite{STAR_DIHADRON_PAPER}, with $corr. = \frac{(SE \ - \alpha ME)}{\alpha ME}$, where $\alpha$ is defined as the ratio of the total number of counts in the SE distribution to the total number of counts in the ME distribution, and is roughly equal to $\frac{1}{N_{ME}}$, with $N_{ME}$ being the total number of mixed events used for each candidate trigger. This correlation quantity is calculated for $D^{0}$ candidates from the signal region of the invariant mass distribution (red band in right panel in Fig. \ref{D0_Decay_Cartoon}), as well as from sidebands (SB) of the invariant mass distribution to estimate the correlations from background K$\pi$ pairs (green bands in the right panel of Fig. \ref{D0_Decay_Cartoon}). Additionally, a correlation is calculated for $D^{0}$-$\pi$ pairs, where the invariant mass of the pair is within the invariant mass region of the charged, excited D-meson state, $D^{*\pm}$. The $D^{*} \rightarrow D^{0} + \pi$ decay happens outside the medium, and is restricted to small opening angle due to the kinematics of the decay. The $\pi$ from this decay adds an associated hadron not originating from medium interactions, thereby increasing the NS associated yield. The final correlation equation including all of these contributions is derived as 

\begin{equation}
\begin{split}
\frac{C_{D^0+h}}{(\alpha ME)_{D^0+h}} & = 
\frac{\rm{S}+\rm{B}}{\rm{S}} \frac{(SE - \alpha ME)_{\rm sig}}{(\alpha ME)_{\rm sig}}  \\
& - \frac{\rm{B}}{\rm{S}} \frac{(SE - \alpha ME)_{\rm SB}}{(\alpha ME)_{\rm SB}} - \frac{\rm{S}+\rm{B}}{\rm{B}} \frac{(\alpha ME)_{D^0 \pi}}{(\alpha ME)_{\rm sig}}
\frac{(SE - \alpha ME)_{D^0 \pi}}{(\alpha ME)_{D^0 \pi}},
\label{MAIN_CORR_EQUATION}
\end{split}
\end{equation}

\noindent{where S and B are the signal and background yields of the $D^{0}$ invariant mass signal region, respectively.}

\section{Results}

\begin{figure}[H]
\centering
\includegraphics[width=4.5 cm]{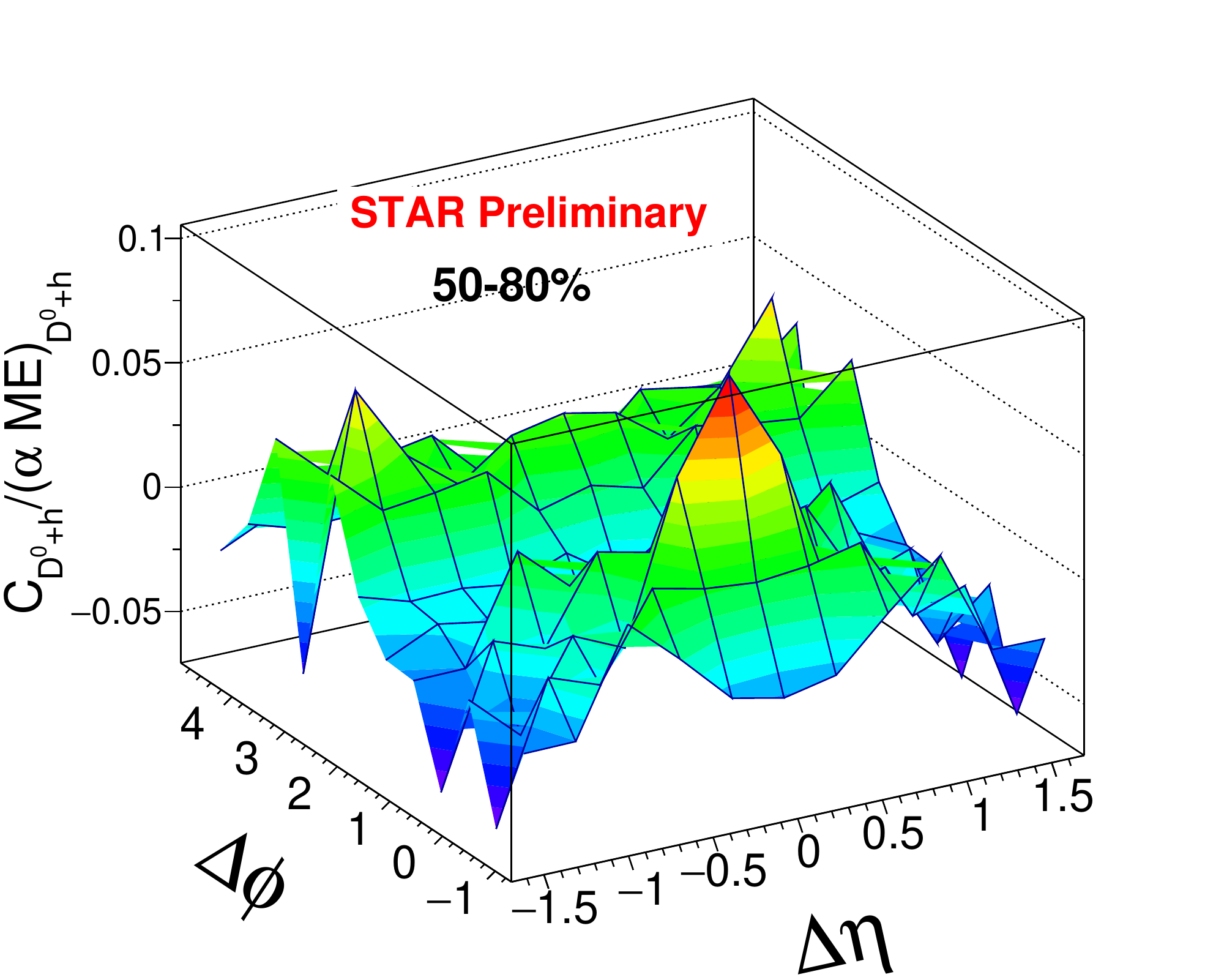}
\includegraphics[width=4.5 cm]{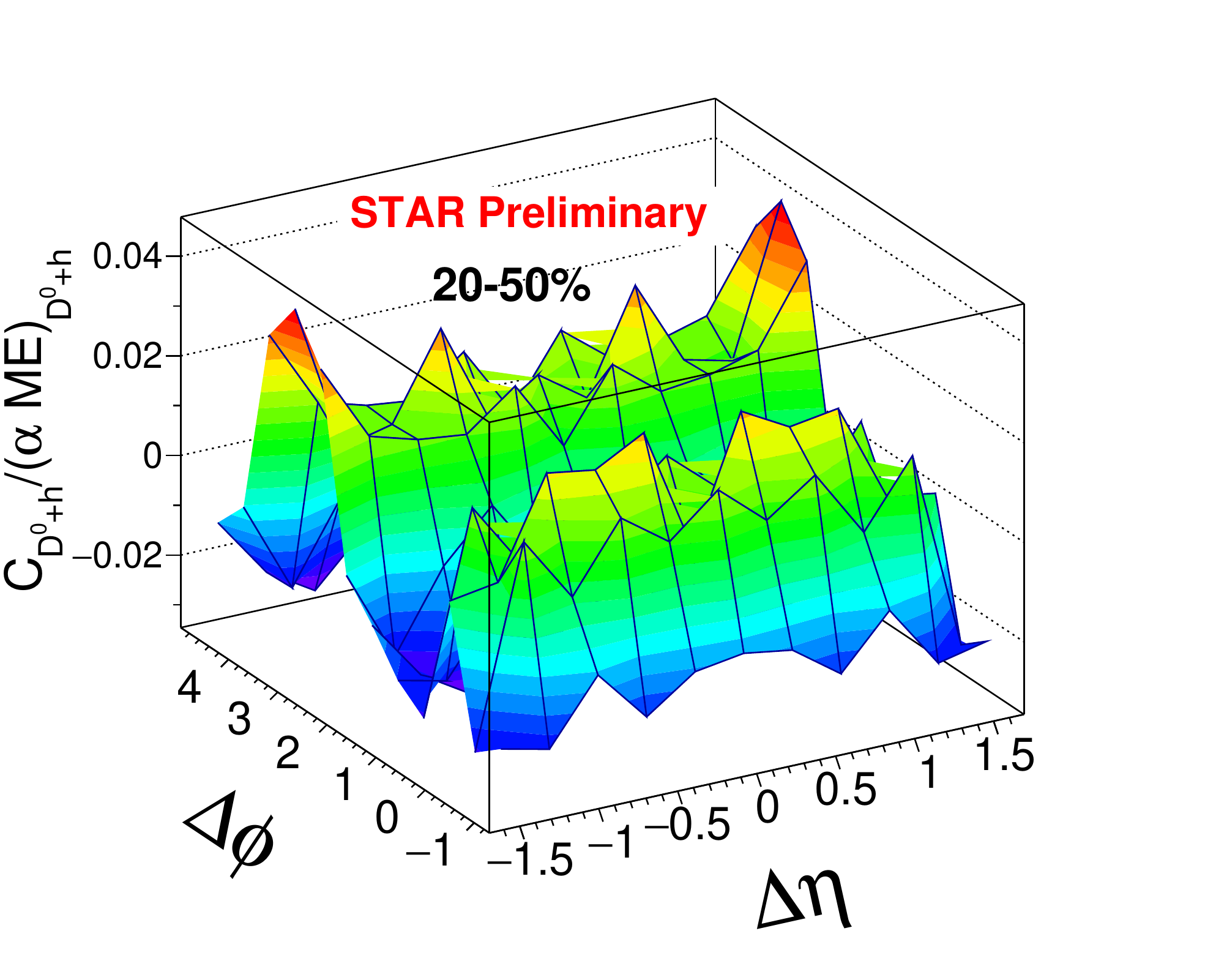}
\includegraphics[width=4.5 cm]{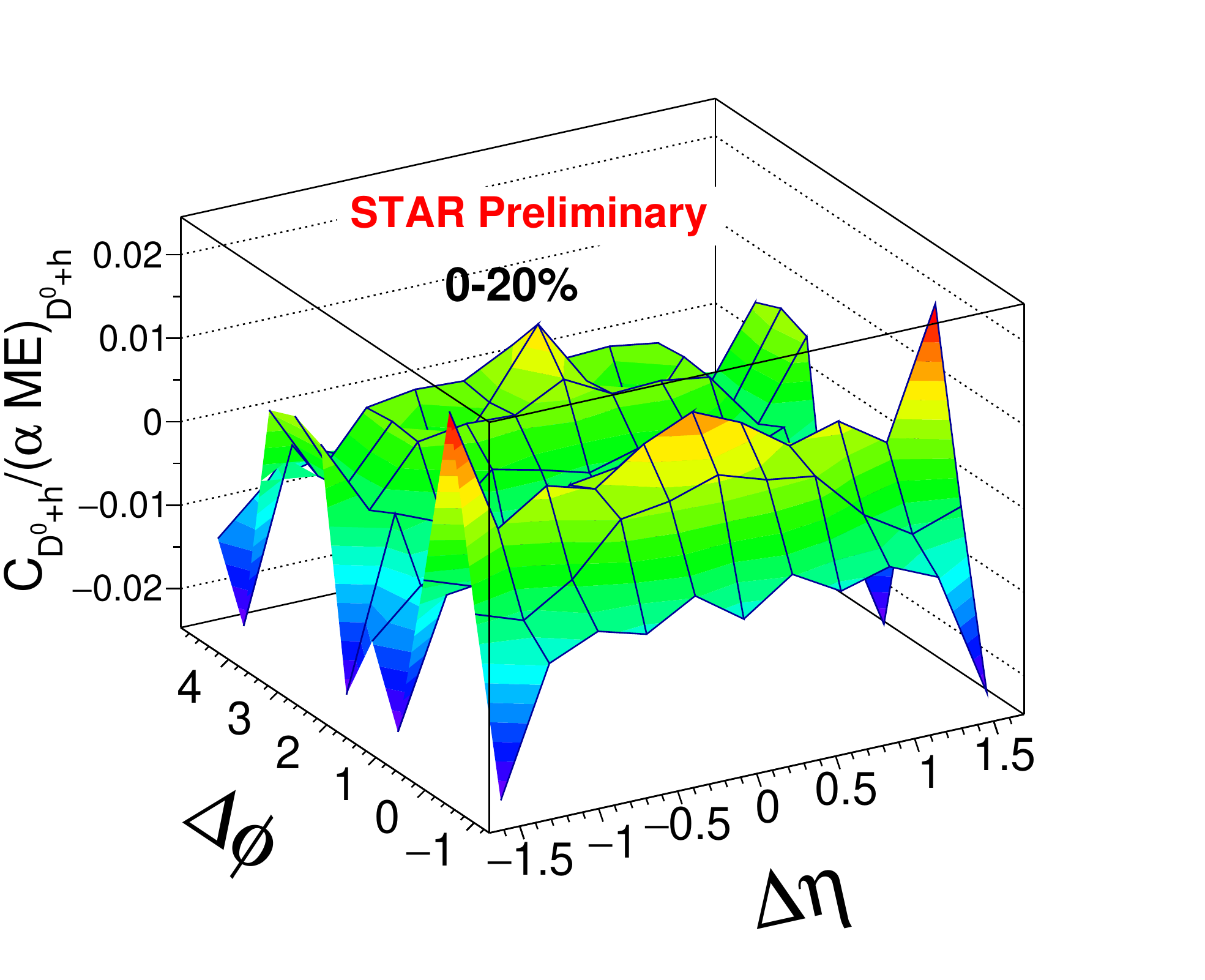}
\caption{$D^{0}$-hadron correlations from Au+Au collisions at $\sqrt{s_{NN}}$ = 200 GeV collisions at STAR. The left-most plot is most-peripheral (50-80\%), the center is mid-central (20-50\%) and the right-most plot is most-central (0-20\%).}
\label{finalCorrelations}
\end{figure}  

Extraction of the underlying physics from $D^0$-hadron correlations, which are shown in Fig. \ref{finalCorrelations}, is carried-out by fitting the data with a multi-parameter model that describes the visible features in the data. We assume a NS 2D Gaussian centered at $(\Delta\eta,\Delta\phi) = (0,0)$, an away-side (AS) 2D Gaussian centered at $(0,\pi)$, a $\Delta\eta$-independent quadrupole, and an overall constant offset. Both 2D Gaussians are required to be periodic on $\Delta\phi$. The model is given by

\begin{equation}
\begin{split}
& F(\Delta\eta,\Delta\phi)  =  A_0 + 2A_{\rm Q} \cos (2\Delta\phi) 
+ A_{\rm NS} e^{-\frac{1}{2} \left[ (\Delta\eta/\sigma_{\Delta\eta,NS})^2 + (\Delta\phi/\sigma_{\Delta\phi,NS})^2 \right]} \\
& + A_{\rm AS} e^{-\frac{1}{2} \left[ (\Delta\eta/\sigma_{\Delta\eta,AS})^2 + ((\Delta\phi - \pi)/\sigma_{\Delta\phi,AS})^2 \right]} + {\rm periodicity},
\label{fitModel}
\end{split}
\end{equation}

\noindent{where near-side Gaussian terms at $\Delta\phi = \pm 2\pi$, etc. and AS Gaussians at $\Delta\phi = -\pi, \pm3\pi$, etc. are not listed but are included in the model to enforce the periodicity condition. Additional and/or alternate model elements, such as a sextupole (realted to $v_3$), were included in the study of systematic uncertainties, but ultimately, the inclusion of additional model elements was limited by the available statistics. Additionally, when the away-side Gaussian width approaches 1, it mathematically limits to a dipole ($\mathrm{cos}(\Delta\phi$)) due to periodicity. This multi-parameter fitting method has been employed previously in \cite{STAR_DIHADRON_PAPER} to describe centrality trends of correlation structures from unidentified hadrons in heavy-ion collisions.}

Using the extracted information from the NS parameters of the fits, the associated yield of hadrons per trigger $D^{0}$ is given by 

\begin{equation}
\begin{split}
Y_{\rm NS,peak}/N_{D^0} = \frac{\mathrm{d}N_{\rm ch}}{2\pi d\eta} \left( 1 - \frac{1}{2N_{\Delta\eta}} \right) 
\int_{\Delta\eta \, {\rm accep}} \hspace{-0.3in} \mathrm{d}\Delta\eta \int \mathrm{d}\Delta\phi F_{\rm NS-peak}
(\Delta\eta,\Delta\phi),
\end{split}
\label{yieldEquation}
\end{equation}

\begin{figure}[H]
\centering
\includegraphics[width=4.5 cm]{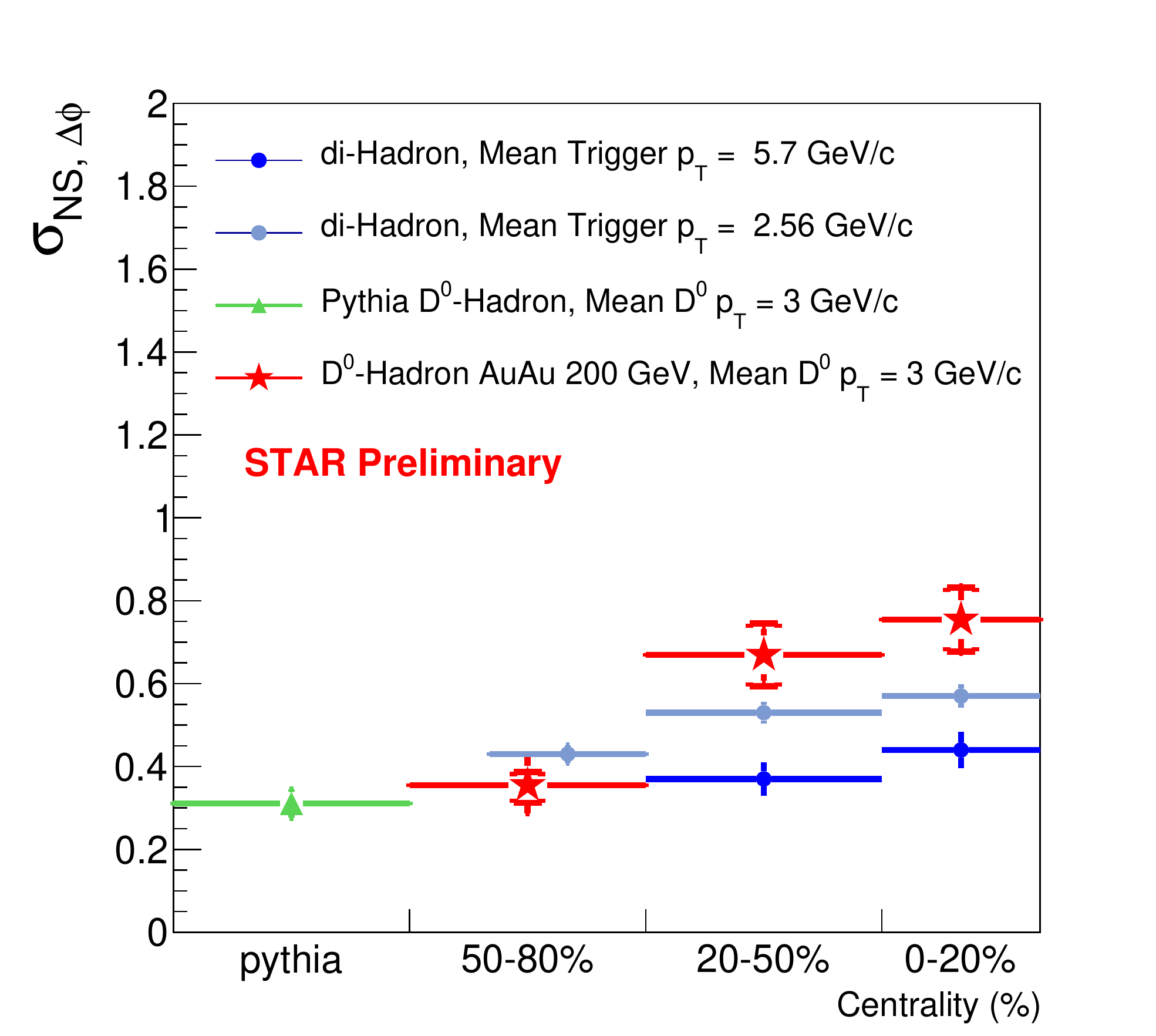}
\includegraphics[width=4.5 cm]{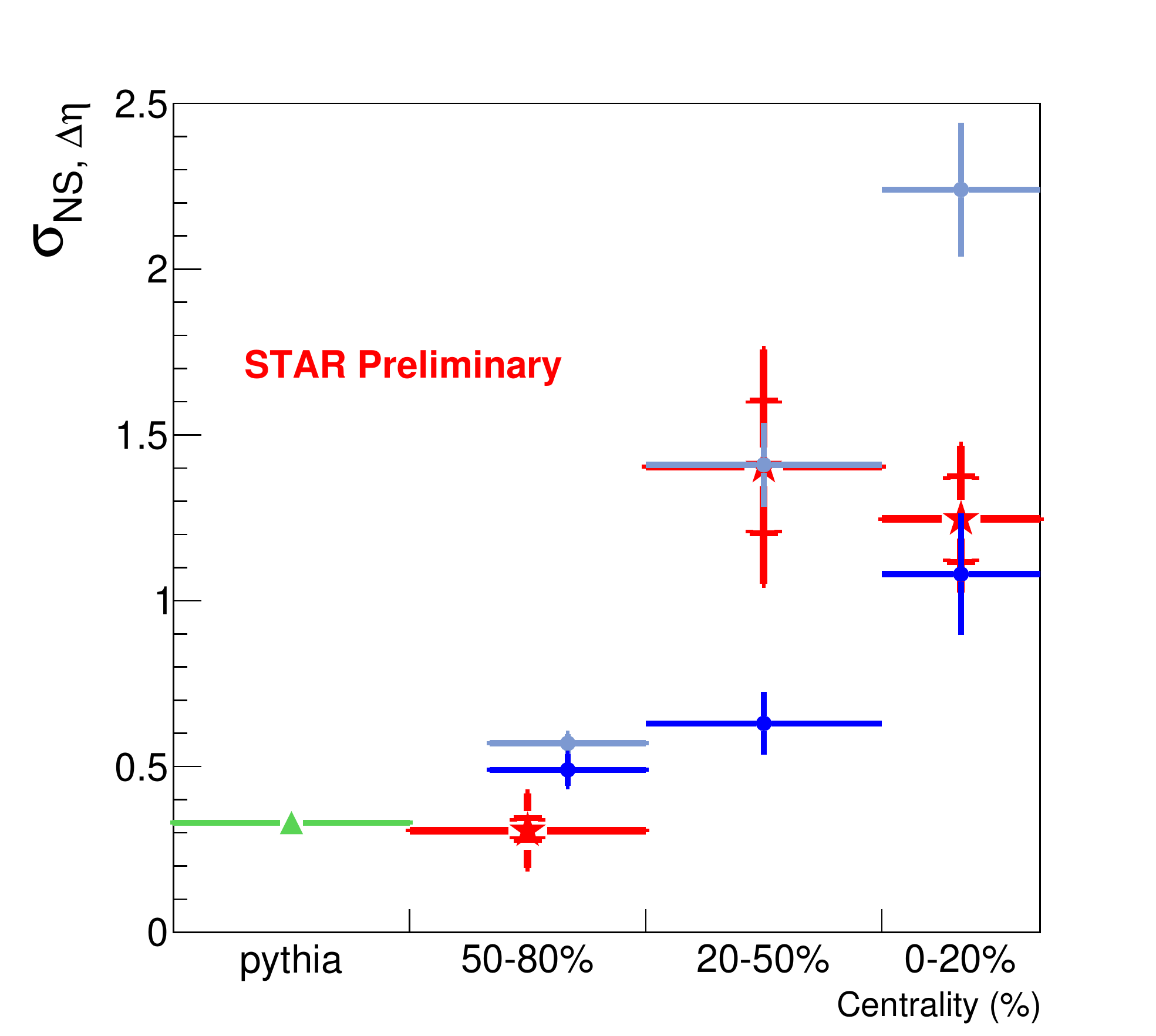}
\includegraphics[width=4.5 cm]{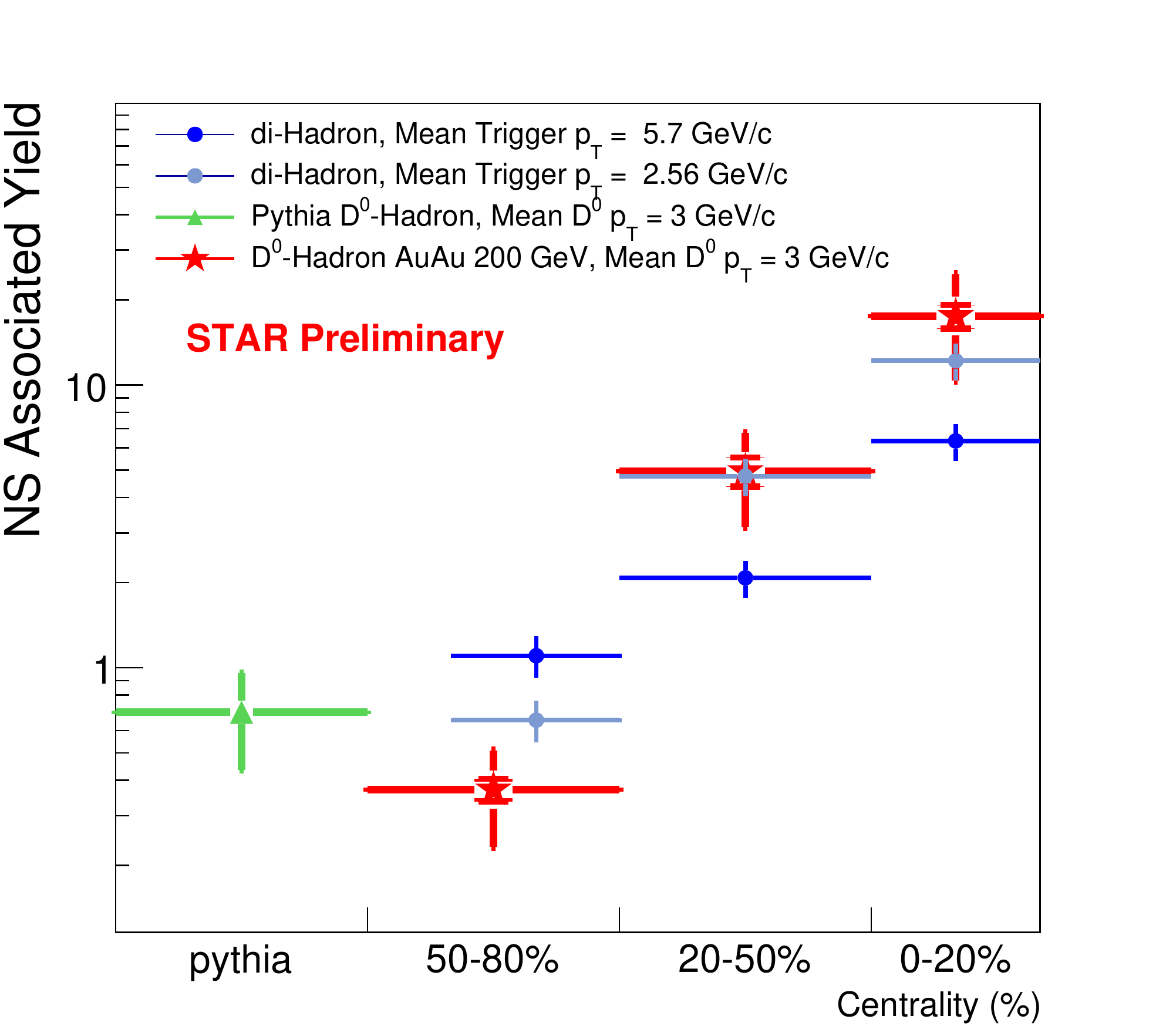}
\caption{Extracted NS widths on $\Delta\eta$ and $\Delta\phi$ and associated yields from $D^{0}$-hadron correlations (red stars). The light and dark blue data points are from \cite{KETTLER_PAPER} and are LF di-hadron correlations with mean $p_{T}$ of 2.56 and 5.7 GeV/$c$, respectively. The green data points are from correlations calculated using PYTHIA 8.23 (tune: \cite{PYTHIA_TUNE_PAPER}).  }
\label{widthAndYieldPlots}
\end{figure}

\noindent{where $\frac{\mathrm{d}N_{\rm ch}}{2\pi \mathrm{d}\eta}$ is calculated from \cite{STAR_DIHADRON_PAPER}, and the integrals are over the 2D Gaussian, NS peak using the fit parameters from the fitting procedure. The results of the fitting and calculation of the associated yield are shown in Fig. \ref{widthAndYieldPlots}.}

\section{Discussion and Conclusions}

Presented here is the first measurement of $D^{0}$-hadron correlations as a function of centrality on ($\Delta\eta, \Delta\phi$) in heavy-ion collisions. Two-particle correlations using a $D^{0}$ meson as a trigger allow for the study of jet-like phenomena involving charm quarks by studying the evolution of the widths of the near-side peak on $(\Delta\eta, \Delta\phi)$ and the NS associated yield per trigger as a function of centrality. The widths of the jet-like peak broaden from peripheral to central collisions, exhibiting behavior similar to what is seen in correlations with LF mesons at a similar mean-$p_{T}$. The NS yield increases by an order of magnitude from peripheral to central collisions, indicating significant interactions of the charm quark with the QGP. The widths and yields of $D^{0}$-hadron correlations calculated with PYTHIA are consistent with what is measured in the peripheral centrality bin of the present analysis, indicating minimal interactions of the charm-containing jet with the QGP in peripheral heavy-ion collisions. 

\section*{Acknowledgments}
\noindent{This research was funded by the U.S. Department of Energy under grants No. DE-FG02-94ER40845 and No. de-sc0013391.}


\begin{thebibliography}{99}

\bibitem[J. Adams et al. (STAR Collaboration)]{STAR_HADRON_RAA_PAPER}
J. Adams et al. (STAR Collaboration), {\em Physical Review Letters} {\bf 2003}, {\em 91}, 172302

\bibitem[L. Adamczyk et al. (STAR Collaboration)]{STAR_D0_RAA_PAPER}
L. Adamczyk et al. (STAR Collaboration), {\em Physical Review Letters} {\bf 2014}, {\em 113}, 142301 

\bibitem[S. Acharya et al. (ALICE Collaboration)]{ALICE_D_MESON_RAA_PAPER}
S. Acharya et al. (ALICE Collaboration), arXiv: 1804.09083 (2018)

\bibitem[L. Adamczyk et al. (STAR Collaboration)]{STAR_D0_V2_PAPER}
L. Adamczyk et al. (STAR Collaboration), {\em Physical Review Letters} {\bf 2017}, {\em 118}, 212301

\bibitem[K. Ackermann, N. Adams et al.]{STAR_DETECTOR}
K. Ackermann, N. Adams et al., {\em  Nucl. Inst. Meth. A} {\bf 2003}, {\em 499}, Pgs. 624-632

\bibitem[M. Anderson, J. Berkovitz et al.]{STAR_TPC} 
M. Anderson, J. Berkovitz et al., {\em Nucl. Inst. Meth. A} {\bf 2003}, {\em 499}, Pgs. 659-687

\bibitem[G. Contin, L. Greiner, et al.]{STAR_HFT} 
G. Contin, L. Greiner, et al., {\em Nucl. Inst. Meth. A} {\bf 2018}, {\em 907}, Pgs. 60-80

\bibitem[G. Agakishiev et al. (STAR Collaboration)]{STAR_DIHADRON_PAPER}
G. Agakishiev et al. (STAR Collaboration), {\em Physical Review C} {\bf 2012}, {\em 86}, 064902

\bibitem[D. Kettler, T. Trainor]{KETTLER_PAPER} 
D. Kettler, T. Trainor, {\em Physical Review C} {\bf 2015}, {\em 91}, 064910

\bibitem[S. Shi, X. Dong, M. Mustafa]{PYTHIA_TUNE_PAPER} 
S. Shi, X. Dong, M. Mustafa, arXiv:1507.00614 (2015)

\end{thebibliography}
\end{document}